# Strong charge density wave fluctuation and sliding state in PdTeI with quasi-1D PdTe chains

*Hechang Lei,*[1,2,3,4,*] *Kai Liu,*[1,2] *Jun-ichi Yamaura,*[4] *Sachiko Maki,*[4] *Youichi Murakami,*[5] *Zhong-Yi Lu,*[1,2] *and Hideo Hosono,*[3,4,*]

[1]Department of Physics, Renmin University of China, Beijing 100872, China

[2]Beijing Key Laboratory of Opto-electronic Functional Materials and Micro-nano Devices, Renmin University of China, Beijing 100872, China

[3]Materials and Structures Laboratory, Tokyo Institute of Technology, Yokohama 226-8503, Japan

[4]Materials Research Center for Element Strategy, Tokyo Institute of Technology, Yokohama 226-8503, Japan

[5]Institute of Materials Structure Science, High Energy Accelerator Research Organization (KEK), Tsukuba, Ibaraki 305-0801, Japan

In quasi-one-dimensional (quasi-1D) system, the charge density wave (CDW) transition temperature $T_{CDW}$ is usually lower than the mean-field-theory predicted $T_{MF}$ and a CDW fluctuation region exists between them. Here, we investigate the physical properties of PdTeI single crystal containing quasi-1D PdTe chains. Surprisingly, we find that the carrier concentration decreases gradually before the long-range CDW ordering state occurring at $T_1 \sim 110$ K, reflecting the existence of strong CDW fluctuation with possible pseudogap state at $T \gg T_1$ because of dynamic charge separation of Pd ions ($Pd^{3+} \rightarrow Pd^{2+} + Pd^{4+}$). Moreover, the sliding CDW state appears below $T_2 \sim 6$ K. Combined such low $T_2$ with the feature of multiple quasi-1D bands, PdTeI exhibits exotic crossover behavior from negative to huge positive magnetoresistance under magnetic field and field-induced localization. Thus, PdTeI provides a novel platform for studying the CDW fluctuation and the interplay between magnetic field and CDW state.

## I. Introduction

Electron correlations and reduced dimensionality both are central topics in the area of condensed matter physics. Interacting electrons in reduced dimensions tend to form various novel ground states, such as high-temperature superconductors and charge density wave (CDW) states. For strictly one-dimensional (1D) systems, however, the thermal fluctuations are too strong to form long-range CDW states at finite temperature and only short-range correlations are possible [1, 2]. In practice, 1D chains are embedded in a 3D structure with certain interchain coupling. Such systems are quasi-1D rather than purely 1D. Accompanying with fluctuations that are strong in low-dimensional materials, even weak interchain coupling can lead to long-range CDW ordering state and opening a gap at a finite temperature $T_{CDW}$, which is much lower than the mean-field transition temperature $T_{MF}$. Moreover, there is a CDW fluctuation region with the pseudogap state that the CDW energy gap still persists but single-particle coherence disappears between $T_{CDW}$ and $T_{MF}$ [3 – 5]. Although the CDW fluctuation and pseudogap state at $T > T_{CDW}$ have been observed by angle-resolved photoelectron spectroscopy (ARPES), optical conductivity, and tunneling spectroscopy techniques in several quasi-1D CDW materials, such as $K_{0.3}MoO_3$, $NbSe_3$, and organic conductor TTF-TCNQ [3, 6 – 8], the pseudogap state seems to have barely influence on the electronic transport properties like dc resistivity and Hall coefficient as well as corresponding carrier concentration when $T > T_{CDW}$ [9 – 11]. This is ascribed to the highly mobile feature of CDW cluster in the fluctuation regime and thus electronic transport properties behave as if there is no CDW fluctuation in electronic transport measurements [7].

Recent X-ray diffraction (XRD) and neutron powder diffraction (NPD) studies on PdTeI polycrystal with quasi-1D PdTe chains along the *c* axis revealed that there is a dynamic charge separation of Pd ions with local $Pd^{2+}$ and $Pd^{4+}$ pair persisting at high temperature and this dynamic charge disproportionation becomes static without long-range CDW ordering at $T < 50$ K [12]. Moreover, a theoretical study also suggested that this system is susceptible to the instability of charge density [13]. These studies motivates us to examine on the crystal structure, electronic transport properties, and electronic structure of PdTeI single crystals. Surprisingly, there is a significant decrease of carriers concentration when temperature is above the first CDW transition temperature ($T_1 \sim 110$ K), ascribing to the strong CDW fluctuation with possible pseudogap state at $T \gg T_1$. Moreover, there is another anomaly appearing at $T_2$ ($\sim 6$ K) due to the pinning of CDW state possibly induced by incommensurate to commensurate lock-in transition and the sliding CDW state is observed below $T_2$. Because of the low $T_2$ and the feature of multiple quasi-1D bands, PdTeI exhibits unusual electronic transport properties, such as the crossover behavior from negative to huge positive magnetoresistance (MR) under magnetic field.

## II. Methods

High-quality single crystals of PdTeI were grown using hydrothermal method [13, 14]. Pd powder, Te and $I_2$ chunks were mixed together with the ratio of 1 : 1 : 2. The mixture and KI (30 mg) were put in the 55 % HI solution (1 ml) and sealed in a glass ampoule. The ampoule was slowly heated to 300 °C and then kept at this temperature for 1 month. After reaction, the solution was filtered off and the residues were washed by distilled water for several times and dried in the air naturally. Bar-like single crystals with metallic luster up to $0.2 \times 0.2 \times 1$ mm$^3$ can be picked up from residues. The XRD pattern of a PdTeI single crystal was also collected by using a Bruker D8 ADVANCE diffractometer with Cu $K_\alpha$ radiation at room temperature. Synchrotron radiation XRD measurements were performed using a large-curved imaging plate (Rigaku R-AXIS) for searching the weak additional reflection at low temperature. High-resistivity GaAs wafer ($R_s > 10^6$ Ω/square) was employed as a stage to measure resistivity of PdTeI single crystals. Tine Ti (10 nm) - Au (50 nm) electrodes were attached using sputtering method on the GaAs wafer. The thin plates of PdTeI single crystals were cut using the focused ion beam (FIB) technique. These thin plates were adhered to the GaAs wafer using a cryogenic epoxy adhesive (NITOFIX SK-229) and connected to the electrodes using tungsten containing carbon deposited by irradiating Ga ion beam to evaporated $W(CO)_6$. Electrical transport measurements were carried out on a physical property measurement system (PPMS-9, Quantum Design). Longitudinal and transverse resistivity were measured using four-probe configuration with constant current low frequency ac mode ($I = 0.3$ mA and $f = 17$ Hz). Hall resistivity was extracted from the difference of transverse resistance measured at the positive and negative fields, i.e., $\rho_{ca}(H) = [\rho(+H) - \rho(-H)]/2$, which can effectively eliminate the longitudinal resistivity component due to the misalignment of voltage probes. With a four-probe configuration, $I(V)$ curves were measured by current biasing and determining the generated voltage across two voltage probes. First-principles electronic-structure calculations were carried out with the projector augmented wave method [15, 16] as implemented in the VASP software [17 – 19]. The generalized gradient approximation of Perdew-Burke-Ernzerhof [20] for the exchange-correlation potential was adopted. The previous study showed that it is essential to include the van der Waals (vdW) correction when describing PdTeI [21], where here we used the vdW-optB86b functional [22]. The kinetic energy cutoff of the plane-wave basis was set to be 350 eV. A $4\times4\times6$ K-point mesh was employed for the Brillouin zone sampling. The Gaussian smearing with a width of 0.05 eV was used around the Fermi level. Both cell parameters and internal atomic positions were allowed to relax until all forces were smaller than 0.01 eV/Å. The maximally localized Wannier functions method [23] was utilized to analyze the Fermi surfaces.

## III. Structure of PdTeI

As shown in Fig. 1(a), the basic units of PdTeI (space group P4$_2$/*mmc*) are the octahedra of PdTe$_4$I$_2$ and they connect each other by sharing the Te-Te edges, forming quasi-1D channels of PdTe along the *c* axis. For each channel, there are four PdTe chains but they are not parallel to *c* axis straightly (the angle of Te-Pd-Te is about 168.6°) because of the tilt of the octahedron of PdTe$_4$I$_2$. On the other hand, the channels of PdTe are connected mutually by the I-I edges of the octahedron of PdTe$_4$I$_2$ along the *a* and *b* axes. Such connection of PdTe channels forms large empty square channels surrounded by iodine ions along the *c* axis. The XRD pattern taken on a needle shape surface of the single crystal shows only (l00) peaks, indicating that the surface of PdTeI single crystal is parallel to the *bc* (or *ac*) plane and perpendicular to the *a* (or *b*) axis (Fig. 1(b)). Moreover, the needle-like morphology of PdTeI single crystal suggests that the long axis is the *c*-axial direction (inset of Fig. 1(b)). The crystal habit is consistent with the structure of PdTeI.

## IV. Electronic transport properties

There are two features in the temperature dependence of the electrical resistivity $\rho_{cc}(T)$ curves (Fig. 2(a)). One is a negative slope of $\rho_{cc}(T)$ appearing below about 200 K and a remarkable upturn starting at $T^* \sim 130$ K with a maximum at $T_1 \sim 112$ K. The other one is a sharp upturn below $T_2 \sim 6$ K which is saturated at $T_{sat} \sim 3.5$ K. For the transition at $T_1$, there is a weak hysteresis between temperature warming and cooling (inset (a) of Fig. 2(a)). Similar behaviors were observed in typical CDW materials such as Lu$_5$Rh$_4$Si$_{10}$ and Tl$_x$V$_6$S$_8$, but the corresponding hystereses are more distinct than that in PdTeI [24, 25]. In contrast, there is almost no hysteresis for the transition at $T_2$ (inset (b) of Fig. 2(a)). On the other hand, the absolute value of $\rho_{bb}(T)$ (Fig. 2(b)) is much larger than that of $\rho_{cc}(T)$ and the anisotropy of resistivity $\gamma = \rho_{bb}(T) / \rho_{cc}(T)$ is about 18.3 and 10.5 at 300 K and 2 K, respectively. Such an anisotropy is comparable to that in NbSe$_3$ [26], implying the highly anisotropic electronic structure of PdTeI. Compared to the $\rho_{cc}(T)$, the anomalies at $T_1$ and $T_2$ for $\rho_{bb}(T)$ are extremely weak (insets (a) and (b) of Fig. 2(b)). It means that the transitions at $T_1$ and $T_2$ have very small even negligible effects on the $\rho_{bb}(T)$. Combined with grain boundary effect, this explains the results of PdTeI polycrystal [12] that the resistivity anomaly starting at $T_1$ is much weaker than that of $\rho_{cc}(T)$ and the upturn at $T_2$ is not observed.

Figure 3(a) shows the field dependence of Hall resistivity $\rho_{ca}(\mu_0 H)$ for PdTeI single crystal at various temperatures. $\rho_{ca}(\mu_0 H)$ is negative in the whole temperature region, indicating that the dominant carrier is electron-type. Importantly, $\rho_{ca}(\mu_0 H)$ shows significant temperature dependence and exhibits non-linear behavior at $T < 110$ K (Fig. 3(a)). These features undoubtedly indicate that PdTeI is a multi-band metal. Using linear fitting at low (0 - 0.5 T) and high (5 - 9 T) fields, the temperature dependence of low-field and high-field Hall coefficients $R_H(0)$ and $R_H(\infty)$ ($R_H(\mu_0 H) = \rho_{ca}(\mu_0 H) / \mu_0 H$) are obtained (Fig. 3(b)). Both $R_H(0)$ and $R_H(\infty)$ are almost identical at $T > T^*$, because of good linearity of $\rho_{ca}(H)$ for the whole field range, but their values decrease gradually with decreasing temperature. Then, both start to decrease quickly with different slopes when $T$ is just below $T^*$. The minima of $R_H(0)$ and $R_H(\infty)$ reach at $T = 50$ and 40 K, respectively. With decreasing temperature further, $R_H(0)$ and $R_H(\infty)$ increase and finally they decrease again at $T < 6$ K (inset of Fig. 3(b)).

On the other hand, the $\rho_{cc}(T, \mu_0 H)$ and $\rho_{bb}(T, \mu_0 H)$ show exotic response to magnetic field. Magnetic field has minor effect on resistivity when $T > 50$ K (Figs. 3(c), (e) and insets). Between 6 K $\leq T \leq$ 50 K, however, the $\rho_{cc}(T, \mu_0 H)$ and $\rho_{bb}(T, \mu_0 H)$ become larger and a minimum on $\rho_{cc}(T, \mu_0 H)$ appears and shifts to higher temperature with increasing field. Moreover, the anomaly of $\rho_{cc}(T, \mu_0 H)$ below $T_2$ is suppressed gradually by magnetic field. Correspondingly, the MR is very small at high temperature (Figs. 3(d) and (f)). With further decreasing temperature, the positive MR becomes obvious and for $\rho_{cc}(T, H)$, it reaches as large as ~ 600 % at $T_2$ with $\mu_0H$ = 9 T when compared to the MR of $\rho_{bb}(T, \mu_0H)$ (~ 45 %) at same temperature and field. When $T < T_2$, although the MR is always positive for $\rho_{bb}(T, \mu_0H)$, the negative MR is emergent for $\rho_{cc}(T, \mu_0H)$ at low-field region.

## V. High-temperature long-range CDW ordering state and low-temperature sliding CDW state

The single crystal XRD of PdTeI confirms the anomaly of resistivity at $T_1$ originates from CDW transition (Fig. 4(a)). There are only Bragg reflections existing at $T \geq 110$ K and the well-defined satellite peaks start to appear at lower temperature (indicated by red arrows), suggesting the emergence of long-range CDW ordering state. This temperature is consistent with the temperature $T_1$ at which the $\rho_{cc}(T)$ has a maximum value and the $\rho_{bb}(T)$ exhibits anomaly, but lower than $T^*$ in the $\rho_{cc}(T)$. The obtained CDW vector is $q_{CDW}$ = (0, 0, 0.396(3)), thus, the long-range CDW ordering state below 110 K is incommensurate. Moreover, the intensity of the satellite peaks increases with decreasing temperature below 110 K (Fig. 4(b)).

Because the $T_2$ where resistivity shows second anomaly is beyond the lower temperature limit of single crystal XRD measurement, current dependence of $\rho_{cc}(T)$ and temperatures as well as magnetic field dependence of current-voltage $I(V)$ curves were studied in order to determine the origin of this anomaly. As shown in Fig. 5(a), the $\rho_{cc}(T)$ below 5 K exhibits remarkable current dependence. With increasing current, although the $T_2$ only decreases slightly, the $T_{sat}$ shifts to lower temperature quickly with smaller saturation value. Similar behavior has been observed in NbSe$_3$ when $T < T_{CDW}$ [27]. Moreover, at $T < T_2$, the $I(V)$ curves are ohmic (linear) when $V$ is smaller than certain threshold voltage $V_{th}$ and the $I(V)$ curves become non-linear with increased slopes for $V \geq V_{th}$, i.e., the differential resistance $R_d = dV / dI$ decreases sharply (Figs. 5(b) and (c)). With increasing temperature, the linear region becomes narrower, i.e., $V_{th}$ decreases. At higher temperatures, a clear threshold field can no longer be well defined. These features are perfectly consistent with the properties of sliding CDW state which have been observed and well-studied in many CDW materials [1, 28]. It indicates that the anomaly of transport properties at $T_2$ should be due to the pinning of CDW state. The CDW state is pinned at low current (voltage) region and starts to slide when current (voltage) is large enough, which induces significant enhancement of conductivity [1, 18]. When $T \geq T_2$, the $I(V)$ curve exhibits ohmic behavior without any nonlinearities in the whole voltage range and the $R_d$ is much smaller than that for $V < V_{th}$ at lower temperature, in agreement with the breakdown of low-temperature CDW state. The almost flat $R_d$ indicates that Joule heating effects is negligible in the present current range.

On the other hand, the $I(V)$ curves also exhibit strong magnetic field dependence (Figs. 5(d) and (e)). At $T = 2$ K, the $V_{th}$ shifts to lower voltage with increasing field and the sharp change of $R_d$ disappears when $\mu_0 H \geq 1.5$ T. But the $I(V)$ curves are still not fully linear in the whole voltage range until at very high field (6 T). It implies that the CDW state is depinned or destroyed by magnetic field gradually. Figure 5(f) summarizes the temperature and magnetic field dependence of the threshold electric field $E_T = V_{th} / d$ ($d$ is the distance between two voltage probes) determined from the sharp onset of nonlinearity in the $I(V)$ curves. The $E_T$ decreases with increasing temperature and field, but the slopes are different. The $E_T$ seems to be a linear function of temperature. In contrast, the $E_T$ decreases quickly when field slightly increases and then the suppression of $V_{th}$ starts to become mild.

## VI. Electronic structure of PdTeI

In order to further understand the origin of the CDW states and unusual electronic transport behaviors in PdTeI, the electronic structure was calculated. As shown in Fig. 5(a), the finite density of states (DOS) at the Fermi energy level ($E_F$) is consistent with the results of electronic transport measurement. The main contributions at $E_F$ come from Pd and Te atoms. The band structure shows that there are eight bands crossing $E_F$ (Fig. 5(b)). Among them, four bands are only remarkably dispersive along the $c^*$ axis (Γ - Z and A - M lines) and another four bands are dispersive along all three axes. Thus, the Fermi surfaces (FSs) can be divided into two classes (Fig. 5(c)). One is the opened planar-like quasi-1D FSs with different interceptions along Γ - Z line (called P-type, Band 1 - 4). The other is the closed 3D ellipsoid-shaped FSs with the longest principle axis along A - M line (called E-type, Band 5 - 8). It can be seen that the quasi-1D P-type FSs have strong potential for nesting along the $c^*$ direction, which should be closely related to the emergence of CDW states in this system. On the other hand, the calculated carrier concentration (Table 1) indicates that most of carriers are contributed from the P-type bands (96.17 %) rather than the E-type bands (3.83 %).

## VII. Discussion

CDW fluctuation and quais-1D electronic structure have significant influences on the longitudinal resistivity of PdTeI single crystal. Since the $q_{CDW}$ is along the $c^*$ axis, the charge disproportionation is along the $c$ axis (the direction of quasi-1D PdTe chains) in real space for the structure with tetragonal symmetry, as observed in PdTeI polycrystal [12]. The higher $T^*$ (~ 130 K) than the $T_1$ (~ 110 K) (Figs. 2 and 4) clearly indicates that there is a crossover from 1D CDW fluctuation along the $c$ axis to three-dimensional (3D) one at $T^*$. In other words, at $T > T^*$ the CDW fluctuation is strong in PdTe chains but the coupling of fluctuations between neighboring chains is weak because the correlation length perpendicular to the chain direction $\xi_{ab}$ is less than the distance between the chains $d_{inter}$. When $T_1 < T \leq T^*$, the $\xi_{ab}$ becomes larger and the interchain coupling of fluctuations is enhanced gradually. The fluctuation has a 3D character. At $T \leq T_1$, the 3D long-range CDW order is formed finally and it results in the appearance of superlattice structure [1]. Actually, this behavior is also reflected in the resistivity measurement on PdTeI polycrystal [12]. The broad maxima in the $\rho$(T) curves for temperature cooling and warming begin at about 120 K and both $\rho$(T) curves merge together at 50 K which is defined as the CDW transition temperature $T_{CDW}$ of PdTeI polycrystal. That the $T_{CDW}$ is lower than $T_1$ and only the short-range static CDW ordering state appears at ~ 50 K for PdTeI polycrystal [12] could be tentatively ascribed to some combination effects of different crystal sizes/shapes and possible different amount of compositional variations as well as antisite disorder originating from similar ionic sizes of $Te^{2+}$ (0.207 nm) and $I^-$ (0.206 nm) between polycrystal and single crystals.

Theoretically, the zero-field longitude conductivity can be expressed as

$$\sigma_{xx} = |e|\sum_i n_i \mu_i^x = |e|^2 \sum_i n_i \tau_i^x / m_i^{*,x} \quad (1)$$

where $n_i$ and $\mu_i^x$ are the carrier concentration and mobility along the $x$ direction of the $i$th band. $\tau_i^x$ and $m_i^{*,x}$ are the relaxation time and effective mass of corresponding bands along the $x$ direction. $\mu_i^x$ and $m_i^{*,x}$ are positive for both electron and hole-type carriers. Due to much larger $n_i$ of P-type bands than those of E-type bands (Table 1) and comparable $m_i^{*,c}$ for both kinds of bands, the $\rho_{cc}$ ($\approx 1 / \sigma_{cc}$) is mainly determined by the P-type bands. Because the CDW states should induce the gapping of the P-type bands, it has strong influences on $\rho_{cc}$ and Hall resistivity $\rho_{ca}$ as shown in Figs. 2 and 3. On the other hand, although the $n_i$ of the E-type bands are small, their $m_i^{*,b}$ should be much smaller than those of P-type bands (Fig. 6(b)). Thus, the E-type bands still have remarkable even dominant influence on the $\rho_{bb}$. Since the E-type bands are intact when CDW states appear, there are no significant changes of $\rho_{bb}$ at the CDW transition temperatures.

The temperature dependence of Hall coefficients clearly reflects the strong CDW fluctuation and the evolution of CDW states in PdTeI single crystal. If the contributions of E-type bands to the $\rho_{cc}$ and $\rho_{ca}$ are ignored and the four P-type bands are simplified into two, according to the two-band model [29]

$$R_\mathrm{H} = \frac{1}{|e|} \frac{(\mathrm{Sgn}(q_1)n_1\mu_1^2+\mathrm{Sgn}(q_2)n_2\mu_2^2)+(\mathrm{Sgn}(q_1)n_1+\mathrm{Sgn}(q_2)n_2)(\mu_1\mu_2)^2(\mu_0 H)^2}{(n_1\mu_1+n_2\mu_2)^2+(\mathrm{Sgn}(q_1)n_1+\mathrm{Sgn}(q_2)n_2)^2(\mu_1\mu_2)^2(\mu_0 H)^2} \quad (2)$$

$$R_\mathrm{H}(0) = \frac{1}{|e|} \frac{\mathrm{Sgn}(q_1)n_1\mu_1^2+\mathrm{Sgn}(q_2)n_2\mu_2^2}{(n_1\mu_1+n_2\mu_2)^2} \quad (3)$$

$$R_\mathrm{H}(\infty) = \frac{1}{|e|} \frac{1}{\mathrm{Sgn}(q_1)n_1+\mathrm{Sgn}(q_2)n_2} = \frac{1}{|e|n_a} \quad (4)$$

$$\Delta\rho(H)/\rho_0 = \frac{n_1\mu_1 n_2\mu_2(\mathrm{Sgn}(q_1)\mu_1-\mathrm{Sgn}(q_2)\mu_2)^2(\mu_0 H)^2}{(n_1\mu_1+n_2\mu_2)^2+(\mathrm{Sgn}(q_1)n_1+\mathrm{Sgn}(q_2)n_2)^2(\mu_1\mu_2)^2(\mu_0 H)^2} \quad (5)$$

$$R_\mathrm{H}(0)/R_\mathrm{H}(\infty) = 1 + \frac{\mathrm{Sgn}(q_1)n_1}{\mathrm{Sgn}(q_2)n_2} \frac{(1-\frac{\mathrm{Sgn}(q_1)\mu_1}{\mathrm{Sgn}(q_2)\mu_2})^2}{(1+\frac{n_1\mu_1}{n_2\mu_2})^2} \quad (6)$$

$$|\sigma_{cc}(0)R_\mathrm{H}(0)| = \left|\mathrm{Sgn}(q_1)\mu_1 + \frac{1}{1+\frac{n_1\mu_1}{n_2\mu_2}}(\mathrm{Sgn}(q_2)\mu_2 - \mathrm{Sgn}(q_1)\mu_1)\right| \quad (7)$$

where $\mathrm{Sgn}(x) = x/|x|$ ($x \neq 0$) and $q_i$ is 1 (-1) when the carrier type of the $i$th band is hole (electron), the linear field-dependence of $\rho_{ca}(\mu_0 H)$ accompanying with small MR at $T > T_1$ suggests four P-type bands should have identical carrier type and similar mobility (eqs. (2 - 5)). As shown in Fig. 7(a) and Table 1, the apparent carrier concentration $n_a$ (the sum of $n_i$) derived from $R_\mathrm{H}(\infty)$ (eq. (4)) at 300 K is very close to the theoretical total carrier concentration $n_{tot}$ that is the sum of electron-type carriers in all of bands, confirming that the carriers in all bands in PdTeI should be electron-type. Moreover, the $n_{tot}$ is corresponding to 1.007 electrons per formula unit, i.e., the chemical formula of PdTeI can be well expressed as $Pd^{4+}Te^{2-}I^-\cdot e^-$.

On the other hand, the strong temperature dependence of $R_\mathrm{H}(\infty)$ when $T > T_1$ indicates that the FSs are gapped gradually in lowering temperature and this results in a depletion of DOS (decrease of $n_a$). Slightly small $n_a$(300 K) comparing with $n_{tot}$ (Table 1) implies that this process even happens at 300 K, much higher than $T_1$ where the CDW state is emergent. Moreover, a huge number of electrons (~ 80 %) have been gapped when $T$ is just above $T_1$. This unique phenomenon is totally different from weak temperature-dependence of $R_\mathrm{H}(T)$ at $T > T_\mathrm{CDW}$ in other known CDW materials, such as quasi-1D $NbSe_3$, quasi-2D $NbSe_2$, or 3D $R_5Ir_4Si_{10}$ (R = rare earth) [9 – 11]. In addition, the decrease of $n_a$ could cause the negative $d\rho_{cc}(T)/dT$ appearing at about 200 K ( $>> T_1$), distinctly different from the normal metal behavior in NbSe$_3$ when $T > T_{CDW}$ [26]. The study on PdTeI polycrystal shows that there is a dynamic charge disproportionation with local Pd$^{2+}$/Pd$^{4+}$ pairs persisting up to room temperature [12]. Furthermore, similar behavior has been observed in NdFeAsO$_{0.82}$F$_{0.18}$, which is ascribed to the pseudogap effect [30]. Therefore, the strong 1D CDW fluctuation along the $c$ axis should be closely related to the continuous reduction of $n_a$ and there might be a pseudogap state opening at $T > T_1$.

When $T < T_1$, the $n_a$ keeps decreasing with smaller slope until about 50 K (Fig. 7(a)) and the $n_a$(50 K) is still larger than the sum of $n_i$ for the E-type FSs (Table 1), suggesting that those FSs should be intact and the P-type FSs should not be fully gapped. On the other hand, the low-field nonlinear behavior of $\rho_{ca}(\mu_0H)$ and the obvious MR of $\rho_{cc}(\mu_0H)$ (Figs. 3(a) and (c)) indicate that the $\mu_i$ of different bands start to exhibit some difference (eqs. (2) and (5)). The different $\mu_i$ will also lead to the increase of $R_H(0)/R_H(\infty)$ (eq. (6), Fig. 3(b)). Unexpectedly, the $n_a$ increases when temperature decreases from 50 K to 6 K. The origin of this phenomenon is unclear at present and need to be studied further. Because the $E_T$ in PdTeI is very low when compared to other CDW materials, it suggests that the pinning source is most likely to be due to the commensurability effect rather than impurities [1, 28]. Thus, the anomaly at $T_2$ possibly originates from the incommensurate to commensurate lock-in transition as observed in some other CDW materials, such as Er$_5$Ir$_4$Si$_{10}$, 1$T$-TaS$_2$, and K$_{0.30}$MoO$_3$ [31 – 33]. This transition could change the FSs further, consistent with the decrease of $n_a$ at $T < T_2$.

The CDW transitions also play an important role in the changes of mobilities with temperature. The $\sigma_{cc}(0)R_H(0)$ that mainly reflects the evolution of mobility and separates the changes of carrier concentration away (eq. (7)) does not exhibit abrupt change when $T$ crosses $T_1$. Fitting using the expression of $\alpha T^\beta$ gives $\beta = 1.28(2)$ and 1.61(1) for $T > T_1$ and $T < T_1$, respectively (inset of Fig. 7(b)). It suggests the CDW transition at $T_1$ changes the scattering processes of parts of bands, leading to the obvious multiband behaviors at $T < T_1$. In sharp contrast to the behavior of $\rho_{cc}(T)$, these simple power-law relations confirm the strong temperature dependence of $n_a$ discussed above. On the other hand, when $T < T_2$, the sharp drop of $\sigma_{cc}(0)R_H(0)$ reflects the pinning of CDW state which leads to the strong decrease of carrier mobilities.

The low-temperature unusual MR behaviors in PdTeI single crystal can be explained by several reasons. The negative MR for $\rho_{cc}$ at low temperatures (Fig. 3) is dominantly ascribed to the sliding or destroy of CDW state by magnetic field (Figs. 5). In general [34, 35], the Zeeman splitting of the bands at $E_F$ will drive the CDW energy gap to zero, i.e., the magnetic field destroys the CDW state as the Pauli spin energy exceeds the CDW condensation energy. Consequently, it causes the negative MR that has been observed in several CDW systems [36, 37]. Actually, the CDW transition at $T_1$ is also suppressed by magnetic field (Fig. 3(c)) but the shift of transition temperature is very small because it is proportional to $(\mu_B H)^2/k_B T_{CDW}(0)$, where $T_{CDW}(0)$ is the zero-field CDW transition temperature. On the other hand, in a quasi-1D system, when magnetic field is perpendicular to the plane of chains, the Lorentz force induces a sinusoidal modulation of the carrier trajectory along the direction of the chains in real-space and the amplitude is inversely proportional to the field [38], i.e., one-dimensionalization induced by magnetic field. For this process, there is a crossover from negative to positive MR with increasing field because impurities have remarkable influences on the localization of the carriers in a pure 1D state formed at high-field region [39, 40]. When compared to $\sigma_{cc}(0)R_H(0)$, the significantly small $\sigma_{cc}(9\ T)R_H(\infty)$ with peak shifting to higher temperature (~ 20 K) (Fig. 7(b)) also implies the field-induced localization process. Therefore, the magnetotransport behaviors of PdTeI at low temperatures originate from the combined effects of the suppression of CDW by magnetic field, the multiband effects and the one-dimensionalization of electronic structure.

## VIII. Conclusion

In summary, PdTeI with quasi-1D PdTe chains along the $c$ axis exhibits the long-range CDW transition at $T_1 \sim 110$ K, originating from the quasi-1D FSs. Moreover, the CDW state is pinned by possible incommensurate to commensurate lock-in transition at $T_2 \sim 6$ K. In particular, the electron-type carriers are gradually removed from $E_F$ at $T \gg T_1$, strongly suggesting the formation of CDW pseudogap state due to the dynamic charge separation of Pd ions. As far as we know, it might be the first time that the CDW fluctuation with the appearance of a possible pseudogap state has a remarkable influence on transport properties of CDW materials, especially on Hall coefficient and corresponding carrier concentration. Moreover, the existence of multiple quasi-1D bands and comparable energy scale of the pinned or commensurate CDW state at $T_2$ with magnetic field energy result in the peculiar low-temperature magnetotransport behaviors. Therefore, PdTeI provides a good opportunity to understand the characteristic of CDW fluctuation and pseudogap state as well as the interplay between magnetic field and CDW state in the low-dimensional electronic system.

## Acknowledgements

We thank S. Fujitsu for help with FIB experiment. This work was supported by the Ministry of Science and Technology of China (973 Projects: 2012CB921701), the Fundamental Research Funds for the Central Universities, and the Research Funds of Renmin University of China (15XNLF06 and 14XNLQ03), and the National Nature Science Foundation of China (Grant No. 11574394 and No. 11190024), the MEXT Element Strategy Initiative to Form Core Research Center, and Collaborative Research Project of Materials and Structures Laboratory, Tokyo Institute of Technology. Synchrotron radiation experiments were performed at BL-8A in KEK-PF (Proposal No. 2013S2-002). Computational resources have been provided by the PLHPC at RUC.

*E-mail: hlei@ruc.edu.cn; hosono@msl.titech.ac.jp

# Figure Captions

**Figure 1** Crystal structure and single crystal of PdTeI. (a) Crystal structure of PdTeI. The blue, golden, and purple balls represent Pd, Te, and I ions. The PdTe$_4$I$_2$ octahedra are highlighted in green. (b) XRD on a needle-like plane of PdTeI single crystal measured at $T$ = 300 K. Inset: photo of typical single crystals of PdTeI.

**Figure 2 Longitudinal resistivity of PdTeI single crystal at zero field.** (a) Temperature dependence of $\rho_{cc}(T)$ for temperature warming and cooling at zero field. Insets: enlarged parts of $\rho_{cc}(T)$ (a) between 100 and 140 K and (b) below 10 K; (c) the image of a PdTeI single crystal cut by FIB technique and contacted for resistivity measurement. (b) Temperature dependence of $\rho_{bb}(T)$ for temperature warming and cooling at zero field. Insets: enlarged parts of $\rho_{bb}(T)$ (a) between 90 and 130 K and (b) below 10 K.

**Figure 3** Hall effect and magnetoresistence (MR) of PdTeI single crystal. (a) Hall resistivity $\rho_{ca}(\mu_0 H)$ vs $\mu_0 H$ at various temperatures for $I$ // $c$, $E$ // $a$ and $H$ // $b$. (b) Temperature dependence of low-field and high-field Hall coefficients $R_H(0)$ and $R_H(\infty)$ determined from linear fits of the $\rho_{ca}(\mu_0 H)$ curves between 0 - 0.5 T and 5 - 9 T, respectively. Inset: enlarged part of $R_H(0)$ and $R_H(\infty)$ for $T \leq 30$ K. (c) Temperature dependence of $\rho_{cc}(T)$ below 50 K at various magnetic fields. (d) MR of $\rho_{cc}(\mu_0 H)$ at different temperatures. (e) Temperature dependence of $\rho_{bb}(T)$ below 50 K at various magnetic fields. (d) MR of $\rho_{bb}(\mu_0 H)$ at different temperatures. Insets of (c) and (e): $\rho_{cc}(T)$ and $\rho_{bb}(T)$ between 2 K and 300 K at 0 and 9 T, respectively.

**Figure 4** High-temperature Long-range charge density wave ordering state in PdTeI

single crystal. (a) Single crystal XRD photographs of PdTeI at 20.4 K. The arrows indicate the satellite reflections with $q_{CDW}$ = (0, 0, 0.396(3)). (b) Temperature dependence of the integrated intensity of the satellite peak. The satellite reflection appears below $T_1$ ~ 110 K. The typical intensity ratio of the satellite reflections to the fundamental reflections is in the order of $10^{-2}$. The red line is guide-to-eye.

**Figure 5** Low-temperature sliding CDW state in PdTeI single crystal. (a) Temperature dependencies of $\rho_{cc}(T)$ for PdTeI single crystal at low-temperature region at various currents. (b) The $I(V)$ curves and (c) corresponding differential resistance $R_d = dV / dI$ at various temperatures. (d) The $I(V)$ curves and (e) corresponding $R_d$ at various magnetic fields when $T$ = 2 K. (f) Temperature and magnetic field dependence of threshold field $E_T$.

**Figure 6** Calculated electronic structure of PdTeI. (a) Density of states, (b) band structure, and (c) Fermi surfaces of PdTeI. The dashed lines in (a) and (b) denote the position of Fermi energy level, $E_F$.

**Figure 7** Carrier concentration and mobility of PdTeI single crystal. (a) Temperature dependence of $n_a$ determined from $-1/R_H(\infty)$ and (b) $|\sigma_{cc}R_H|$ at low-field and high-field limits. Inset of (b) plots the $|\sigma_{cc}(0)R_H(0)|$ data with the fits using $\alpha T^{-\beta}$ at high- and low-temperature regions in log-log scale.

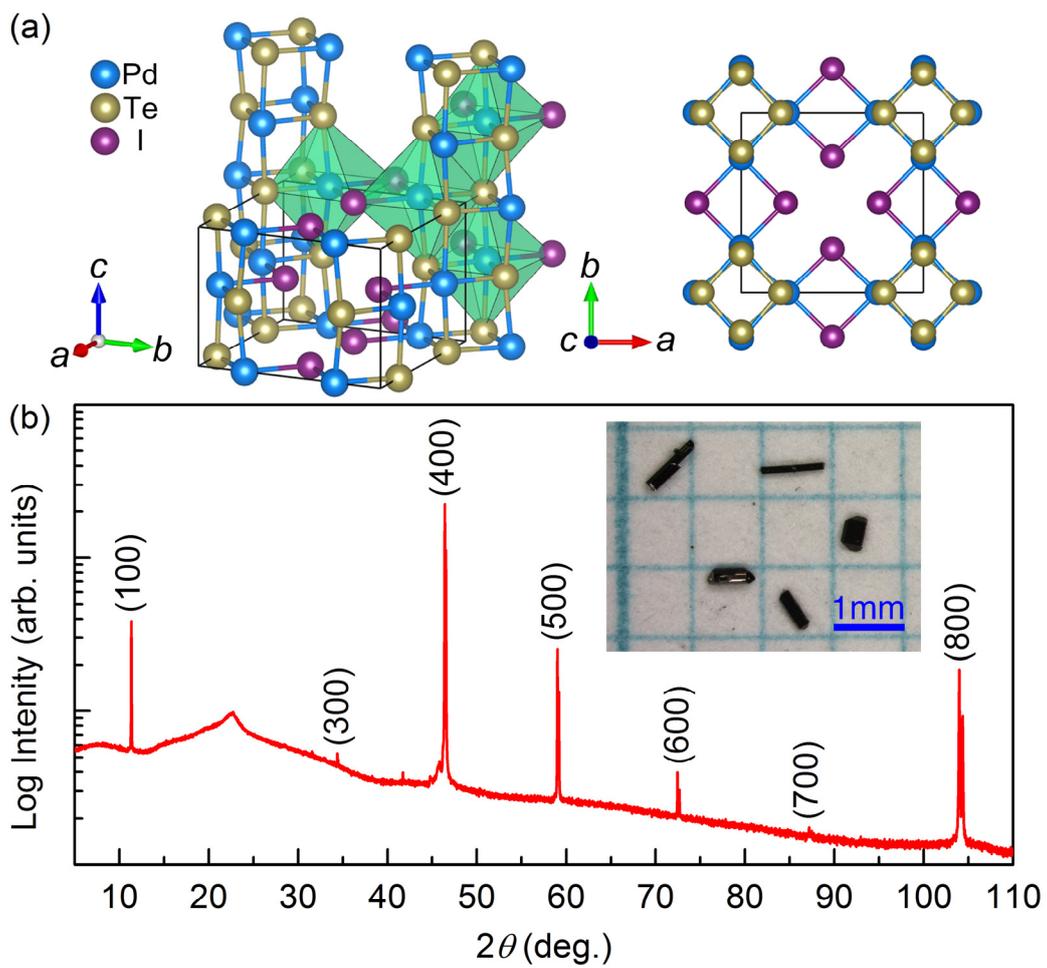

Fig. 1 *Lei* et al.

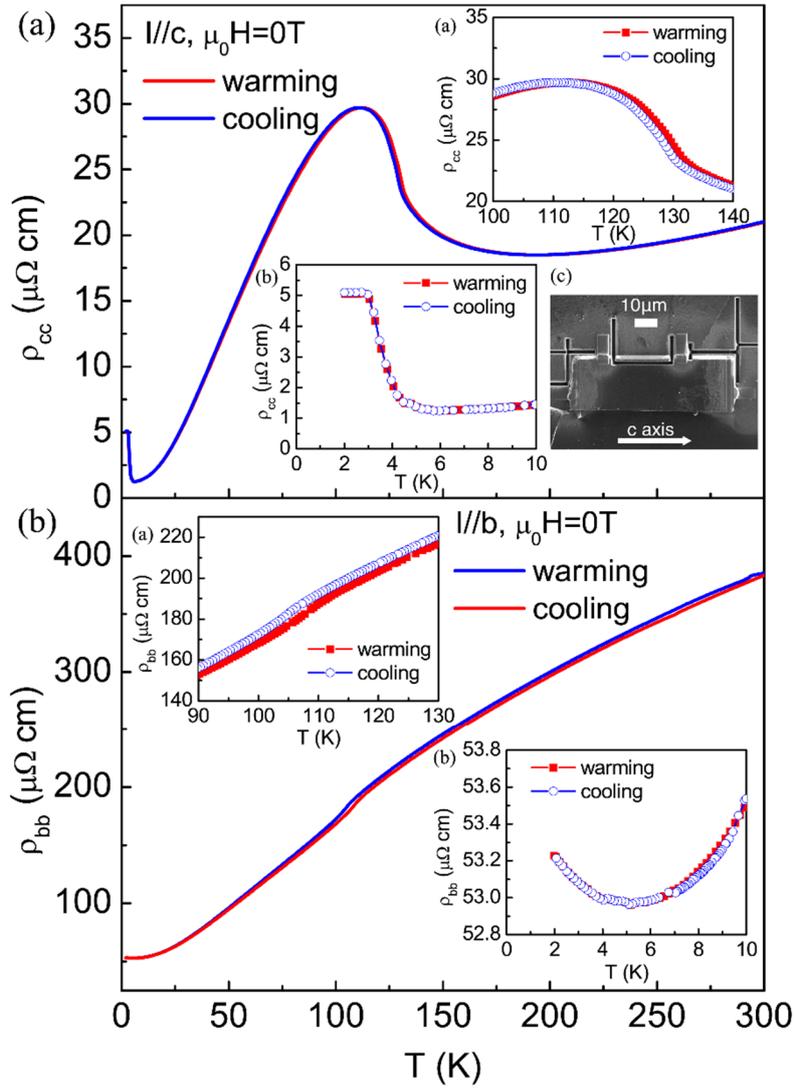

Fig. 2 *Lei* et al.

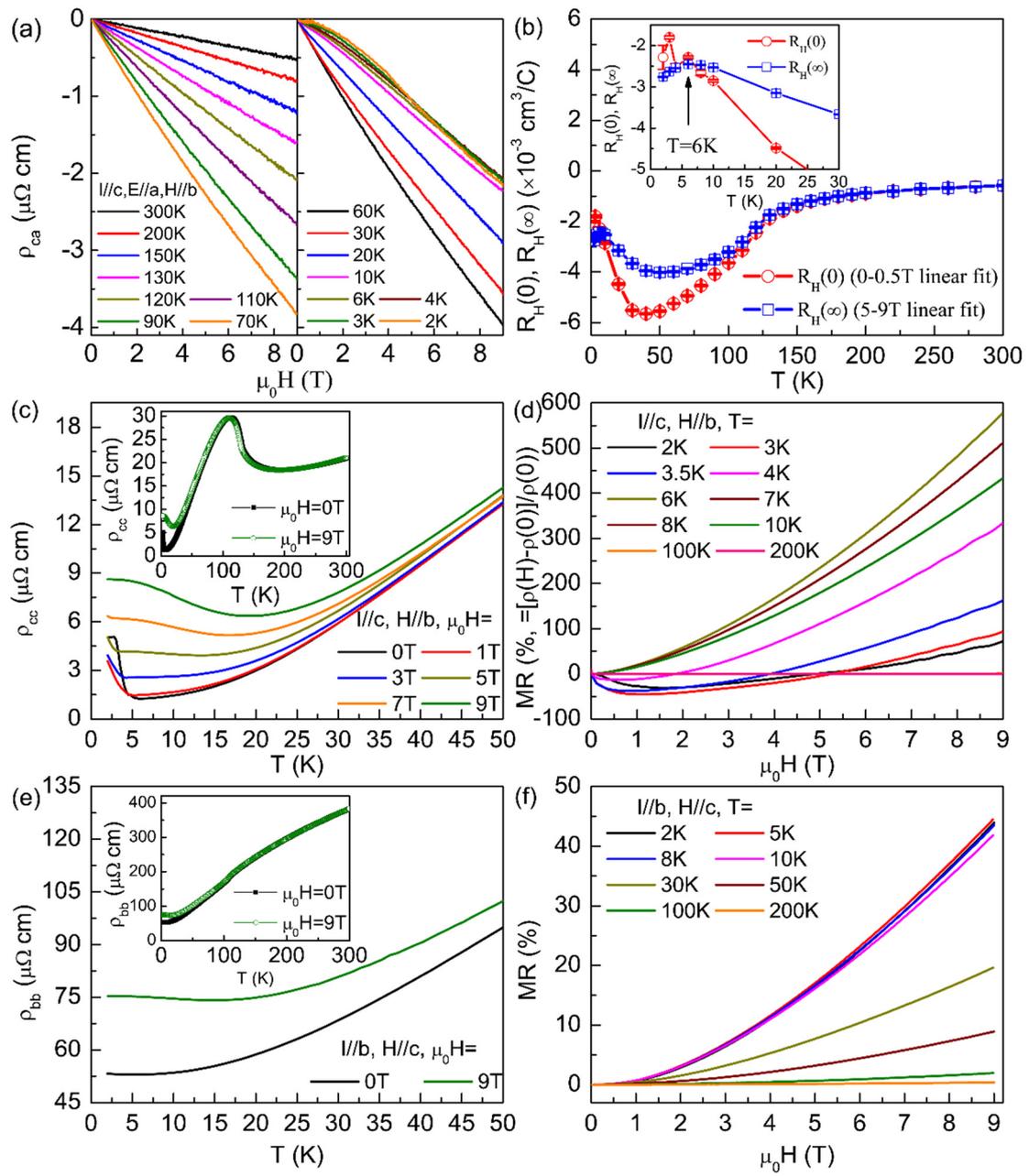

Fig. 3 *Lei* et al.

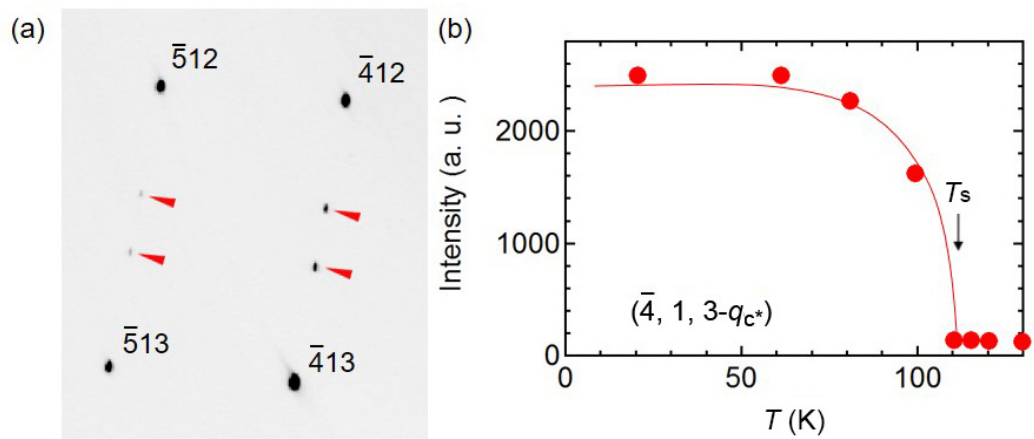

Fig. 4 *Lei* et al.

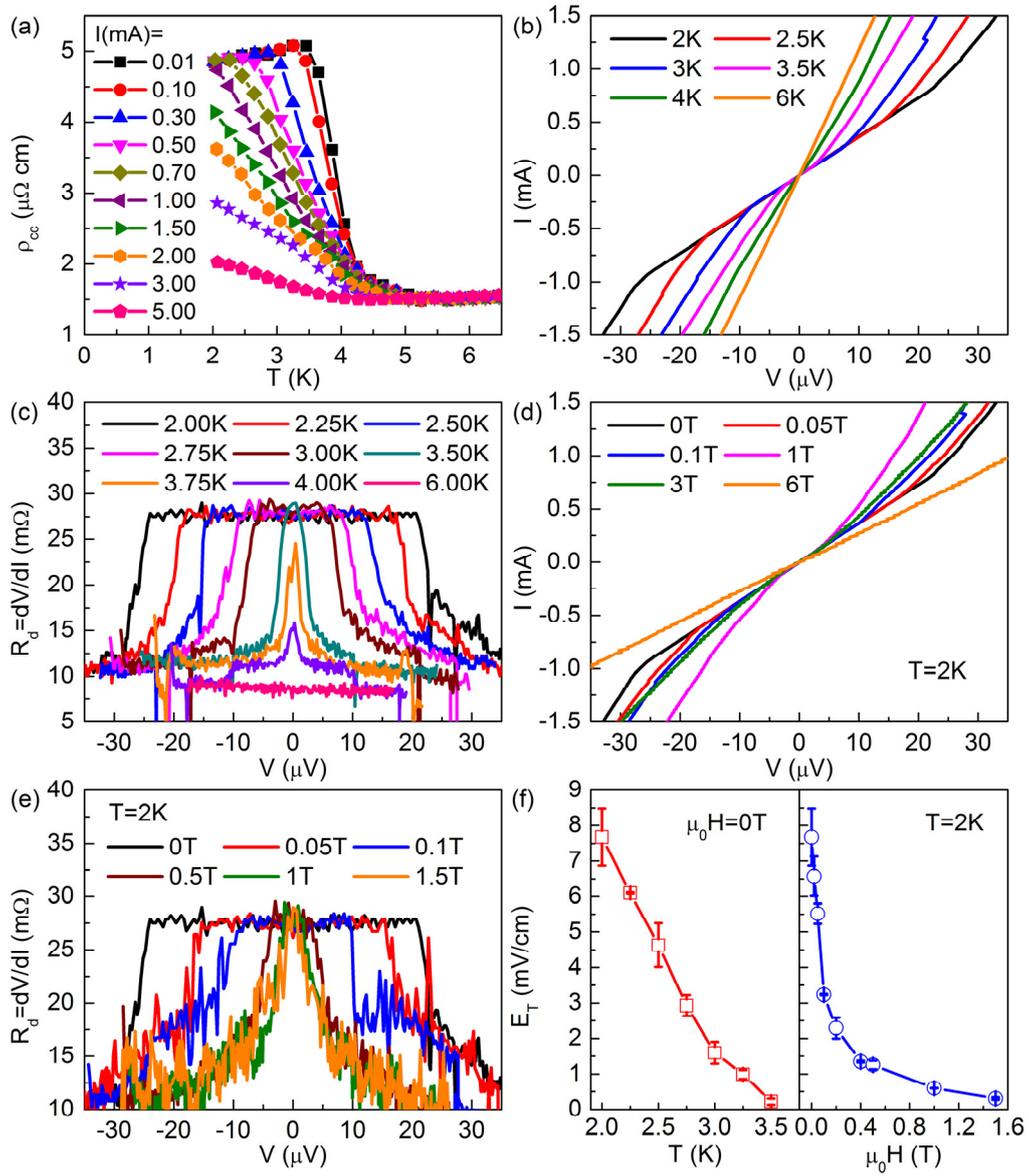

Fig. 5 *Lei* et al.

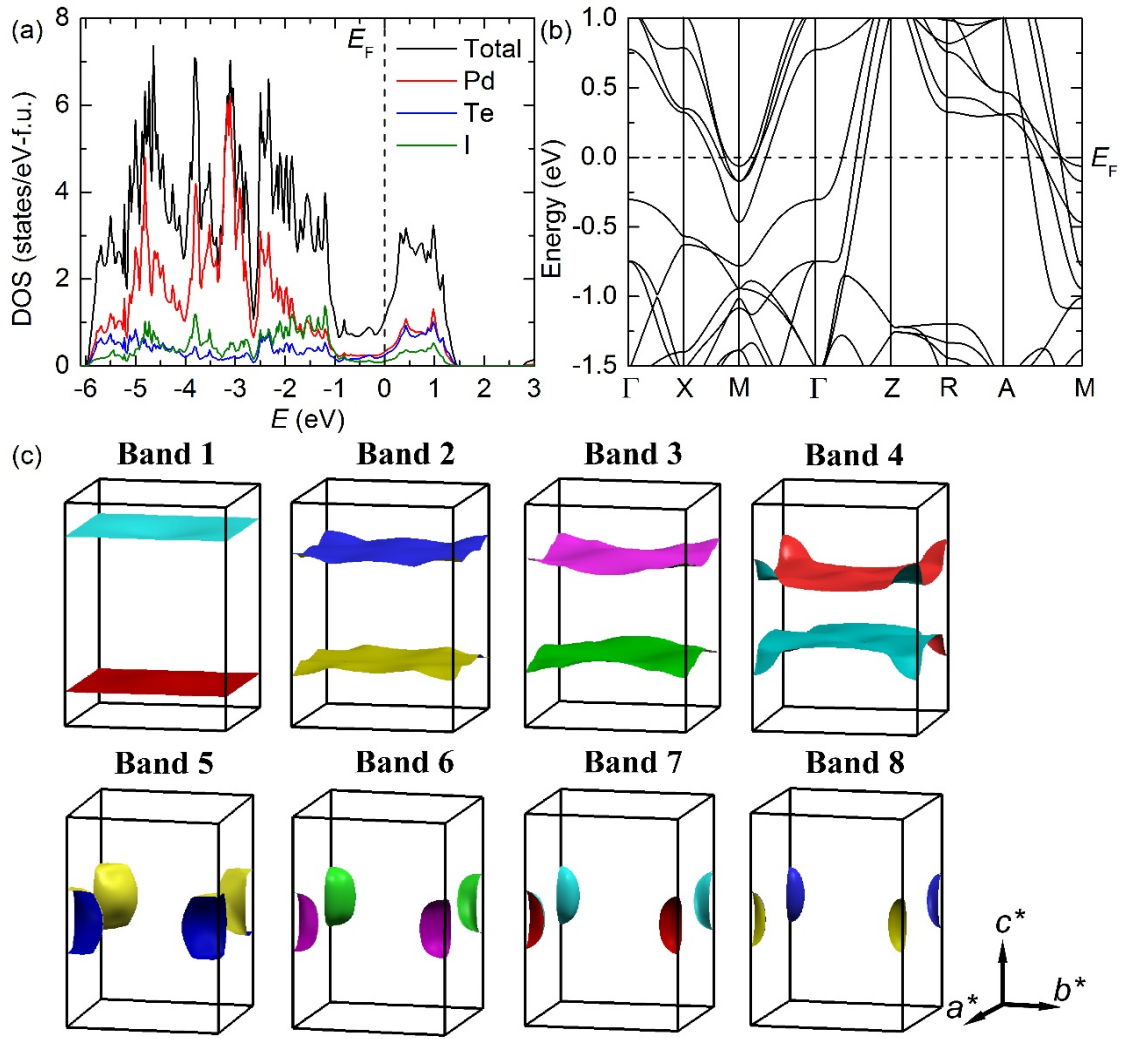

Fig. 6 *Lei* et al.

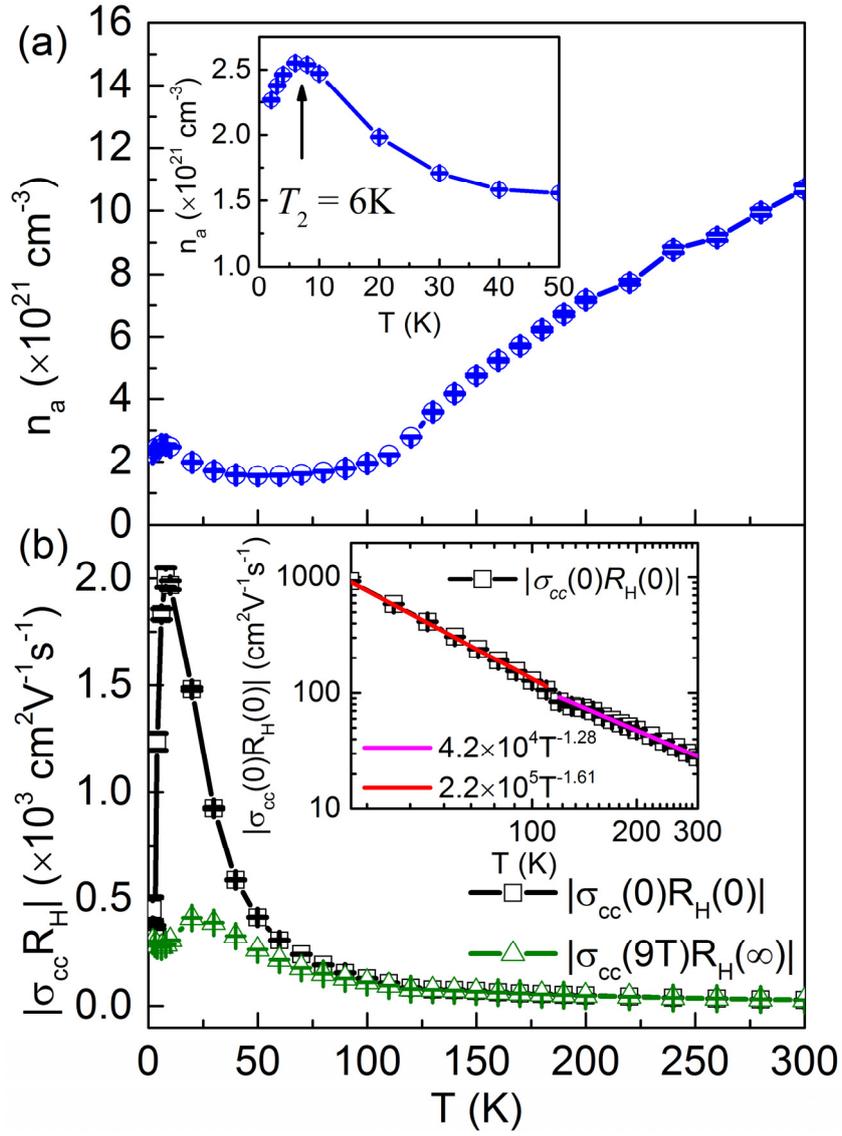

Fig. 7 *Lei* et al.

**Table 1** Theoretical and experimental carrier concentrations of PdTeI. The units of $n_t$, $n_{tot}$ and $n_a$ are $10^{21}$ cm$^{-3}$. The optimized theoretical volume of unit cell is 351.532 Å$^3$ ($Z$ = 4). For one fully occupied band (2 electrons in the band), the $n_{t,max}$ is 5.689 × 10$^{21}$ cm$^{-3}$. The percentages of carrier concentration are calculated by setting $n_{tot}$ as 100 %.

|      | Theory |       |           | Experimental $n_a$ |         |         |         |
| :--: | :----: | :---: | :-------: | :----------------: | :-----: | :-----: | :-----: |
| Band | $n_t$  |   %   | $n_{tot}$ |        300 K       |  110 K  |   50 K  |   6 K   |
|  1   |  3.92  | 34.20 |           |                    |         |         |         |
|  2   |  2.84  | 24.74 |           |                    |         |         |         |
|  3   |  2.56  | 22.29 |           |                    |         |         |         |
|  4   |  1.71  | 14.94 |   11.46   |        10.69       |   2.22  |   1.56  |   2.55  |
|  5   |  0.25  |  2.14 |   100 %   |       93.28 %      | 19.37 % | 13.61 % | 22.25 % |
|  6   |  0.10  |  0.87 |           |                    |         |         |         |
|  7   |  0.06  |  0.55 |           |                    |         |         |         |
|  8   |  0.03  |  0.27 |           |                    |         |         |         |